\documentclass[conference]{IEEEtran}
\IEEEoverridecommandlockouts
\usepackage{amsmath}
\usepackage{graphicx}
\usepackage{amssymb}
\usepackage{hyperref}

\usepackage{tikz}
\usepackage{pgfplots}
\usepackage{subcaption}
\usepackage{xcolor}
\usepackage{booktabs}
\usepackage{array}

\usepackage{tabularray}
\UseTblrLibrary{amsmath,booktabs,counter,diagbox,nameref,siunitx,varwidth,zref}
\usepackage{tikz}

\usepackage{float}

\def\filter{{\mathbf h}}
\def\polynomialMatrix{{\mathbf B}}
\def\fractionalAccuracy{\Delta}

\definecolor{darkpastelpurple}{rgb}{0.59, 0.44, 0.84}
\definecolor{myred}{HTML}{E76F51}
\definecolor{myblue}{HTML}{376996}
\definecolor{mygreen}{HTML}{2A9D8F}
\definecolor{mypurple}{HTML}{822E81}
\definecolor{mywhite}{HTML}{f0dfbb}
\definecolor{darkgray}{rgb}{0.2, 0.2, 0.2}

\def\BibTeX{{\rm B\kern-.05em{\sc i\kern-.025em b}\kern-.08em
    T\kern-.1667em\lower.7ex\hbox{E}\kern-.125emX}}

\begin{document}

\title{Efficient Sub-pixel Motion Compensation in Learned Video Codecs}

\author{
    \IEEEauthorblockN{Théo Ladune, Thomas Leguay, Pierrick Philippe, Gordon Clare, Félix Henry}
    \IEEEauthorblockA{\textit{Orange Research}, France \\ \texttt{theo.ladune@orange.com}}
}
\maketitle

\maketitle

\begin{abstract}
Motion compensation is a key component of video codecs. Conventional codecs
(HEVC and VVC) have carefully refined this coding step, with an important focus
on sub-pixel motion compensation. On the other hand, learned codecs achieve
sub-pixel motion compensation through simple bilinear filtering. This paper
offers to improve learned codec motion compensation by drawing inspiration from
conventional codecs. It is shown that the usage of more advanced interpolation
filters, block-based motion information and finite motion accuracy lead to
better compression performance and lower decoding complexity. Experimental
results are provided on the Cool-chic video codec, where we demonstrate a rate
decrease of more than 10~\% and a lowering of motion-related decoding complexity
from 391 MAC per pixel to 214 MAC per pixel.
\end{abstract}
\section{Introduction}
\label{sec:intro}

Video compression relies on motion compensation to leverage temporal
redundancies between video frames. This includes conventional codecs such as
HEVC \cite{overview-hevc-sullivan}, VVC \cite{overview-vvc-bross} and ECM
\cite{ecm-16-seregin}, autoencoders \textit{e.g.} AIVC \cite{aivc-ladune}, or
the DCVC series \cite{dcvc-hem-li,dcvc-dc-li,dcvc-fm-li} and overfitting-based
codecs like Cool-chic \cite{coolchic-video-3.1-leguay,coolchic-video-4-leguay}.
Motion compensation predicts a video frame through the application of motion
fields to previously received frames, allowing the transmission of only the
prediction error. As such, a more accurate prediction leads to less information
transmitted.

A motion field represents the two-dimensional displacement (motion vector) of
each pixel between video frames. Several motion field characteristics influence
the complexity and quality of the temporal prediction. This paper studies three
of them. \textit{i)} The spatial resolution of the motion field \textit{e.g.}
one motion vector per pixel versus one motion vector for a block of pixels.
\textit{ii)} The precision of the motion vector value \textit{e.g.}, quantized
to quarter pixel accuracy. \textit{iii)} The interpolation filter used to
compute these fractional pixel values. All those factors must be considered
in the light of a performance complexity trade-off, inherent to any practical
application of video compression.
\newline

While conventional codecs have been carefully refining motion compensation
throughout their successive generations, this has been mostly overlooked by the
learned video coding community.  Learned codecs rely on a 2-tap (bilinear)
interpolation filter instead of exploiting more advanced filters as done by
conventional codecs. Moreover, learned codecs use pixel-wise motion vector with
near infinite accuracy compared to the block-based finite-precision motion
vector used \textit{e.g.}, in VVC. This leads learned codecs to have an
increased motion compensation complexity.

This paper studies motion compensation in a learned video coding setting, more
specifically within the low complexity open-source Cool-chic video codec
\cite{coolchic-open-source-orange}. It is shown that a longer interpolation
filter coupled with a quantized and block-based motion field increases the
rate-distortion performance while decreasing the decoding complexity. These
contributions lead to a rate decrease of more than 10~\% and lowers the
motion-related decoded complexity from 391 MAC per pixel to 214 MAC per pixel.
This allows Cool-chic performance to come close to HEVC (+14.3~\% BD-rate) while
having an overall inter-frame decoding complexity of 788 multiplications per
pixel. All contributions are made open-source at
\url{https://github.com/Orange-OpenSource/Cool-Chic}
\cite{coolchic-open-source-orange}.

\begin{figure}[t]
    \centering
    \includegraphics[width=\linewidth]{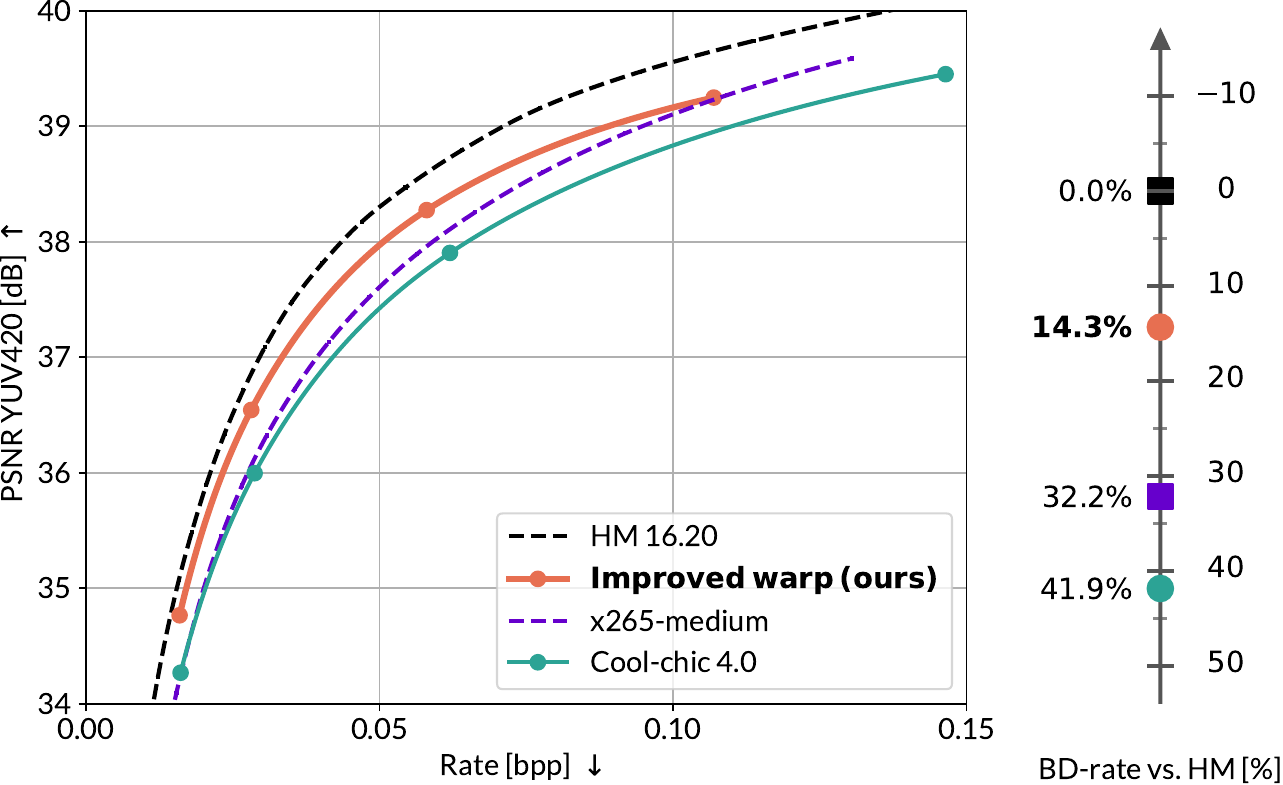}
    \caption{Rate-distortion performance on the first 33 frames from  the JVET
    class B \cite{hevc-ctc-bossen} dataset. Cool-chic 4.0 is from
    \cite{coolchic-video-4-leguay}.}
    \label{fig:rd-curve-jvet_b}
\end{figure}

\section{One dimensional motion compensation}

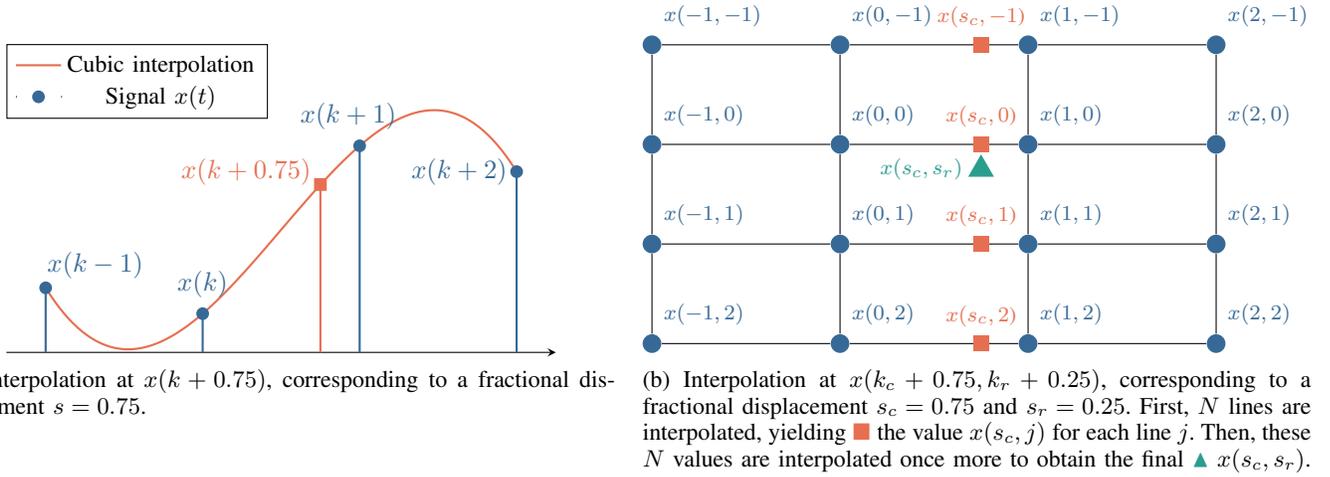
\begin{figure*}
    \centering
    \begin{subfigure}[t]{0.49\linewidth}
        \centering
    \begin{tikzpicture}
        \begin{axis}[
            axis x line=middle,      % Makes axis go through the origin
            axis y line=none,      % No y axis
            xmin=-0.25, xmax=3.25,  % x limits
            ymin=0, ymax=1.99,     % y limits
            height=5cm,
            width=\linewidth,
            legend style={
                at={(0.0,1.20)},
                anchor=north west
            },
            ticks=none,
        ]
            % Plot the estimator
            \addplot[
                domain = 0:3,   % Domain plot
                samples=1000,       % Number of samples points
                myred,              % Color of the curves
                thick,              % Thicker plot
                % forget plot,        % No legend
            ] {-0.5 * (x-1)^3 + 0.75 * (x-1)^2 + 1.05 * (x-1) + 0.3};
            % \node[myred, above] at (axis cs:1.2, 1.1) {Interpolator $f(x)$};

            \addlegendentry{\small Cubic interpolation}

            % Plot the interpolate point
            \addplot[
                ycomb,              % Discrete signals
                myred,              % Color of the plot
                thick,              % Thicker plot
                mark=square*,       % Squared mark
                forget plot,        % No legend
            ] coordinates
            {
                (1.75, 1.3)
            };

            % Print the interpolate point label
            \node[myred, left] at (axis cs:1.75, 1.4) {$x(k + 0.75)$};

            % Plot the discrete know points
            \addplot[
                ycomb,              % Discrete signals
                myblue,             % Color of the plot
                thick,              % Thicker plot
                mark=*,             % Rounded mark,
                % forget plot,        % No legend
            ] coordinates
            {
                (0, 0.5)
                (1 , 0.3)
                (2 , 1.6)
                (3 , 1.4)
            };
            \addlegendentry{\small Signal $x(t)$}

            Print the discrete points label
            \node[myblue, above right, xshift=-0.1cm] at (axis cs:0, 0.5) {$x(k-1)$};
            \node[myblue, above, yshift=0.1cm] at (axis cs:1, 0.3) {$x(k)$};
            \node[myblue, above, yshift=0.1cm, xshift=-0.15cm] at (axis cs:2, 1.6) {$x(k + 1)$};
            \node[myblue, above left, yshift=-0.3cm] at (axis cs:3,1.4) {$x(k + 2)$};

        \end{axis}
    \end{tikzpicture}
    \caption{Interpolation at $x(k + 0.75)$, corresponding to a fractional
    displacement $s = 0.75$.}
    \label{fig:1d-interpol}
    \end{subfigure}
    \hfill
    \begin{subfigure}[t]{0.49\linewidth}
        \usetikzlibrary{shapes.geometric}
        \tikzset{
            bluedot/.style 2 args={fill, circle, myblue, inner sep=2.5pt,
            label={#1:\footnotesize \textcolor{myblue}{#2}}}
        }
        \tikzset{
            reddot/.style 2 args={fill, regular polygon,regular polygon sides=4, myred,
            inner sep=2pt, label={#1:\footnotesize \textcolor{myred}{#2}}}
        }

        \tikzset{
            greendot/.style 2 args={fill, regular polygon,regular polygon sides=3,
            mygreen, inner sep=2pt, label={#1:\footnotesize \textcolor{mygreen}{#2}}}
        }

        \newcommand{\scalex}{2.5}
        \newcommand{\scaley}{1.325}
        \centering
        \begin{tikzpicture}
            \centering
            % Blue, known dots
            \foreach \x in {0,...,3}
                \foreach \y in {0,...,3}
                {
                    \pgfmathtruncatemacro{\rowidx}{-\y + 3 - 1}
                    \pgfmathtruncatemacro{\colidx}{\x - 1}
                    \node (\x\y) [bluedot={75}{$x(\colidx, \rowidx)$}] at (\scalex * \x, \scaley * \y) {};
                }

            \foreach \x in {0,...,3}
                \foreach \y [count=\yi] in {0,...,2}
                    {
                        \draw (\x\y)--(\x\yi) (\y\x)--(\yi\x) ;
                    }

            % Red unknown nodes
            \foreach \x in {1.75}
                \foreach \y in {0,...,3}
                {
                    \pgfmathtruncatemacro{\rowidx}{-\y + 3 - 1}

                    \pgfmathtruncatemacro{\rowidx}{-\y + 3 - 1}
                    \node[reddot={90}{$x(s_c, \rowidx$)}] at (\scalex * \x, \scaley * \y) {};
                }

            % green unknown nodes
            \foreach \x in {1.75}
                \foreach \y in {1.75}
                {
                    \pgfmathtruncatemacro{\rowidx}{-\y + 3 - 1}
                    \node[greendot={180}{$x(s_c,s_r)$}] at (\scalex * \x, \scaley * \y) {};
                }

        \end{tikzpicture}
        \caption{ Interpolation at $x(k_c + 0.75, k_r + 0.25)$, corresponding to
            a fractional displacement $s_c = 0.75$ and $s_r = 0.25$. First,
            $N$ lines are interpolated, yielding
            \textcolor{myred}{$\blacksquare$} the value $x(s_c, j)$ for each
            line $j$. Then, these $N$ values are interpolated once more to
            obtain the final \textcolor{mygreen}{$\blacktriangle$} $x(s_c,
            s_r)$. }
        \label{fig:2d-interpol}
    \end{subfigure}
    \caption{Cubic interpolation of a signal $x$ in one and two dimensions.}

\end{figure*}

This section lays out the mathematical foundations of sub-pixel motion
compensation and presents the role played by the interpolation filter through a
one-dimensional example.

\subsection{Problem statement}

Let $x(t)$ be a one-dimensional signal regularly sampled at $t = kT_s$ with $k
\in \mathbb{Z}$. For simplicity, a unitary sampling period $T_s = 1$ is
considered. Motion compensation or \textit{warping} shifts the input signal by
applying the motion field $v(t)$, which describes the displacement of each point
of $x(t)$. This yields the shifted signal $y(t)$:
\begin{equation}
    y(t) = \mathrm{warp}(x, v)(t) =  x(t + v(t)).
\end{equation}
However, $t + v(t)$ is not necessarily a multiple of the sampling period. As
such, $x(t + v(t))$ must be interpolated from neighboring (known) values to
handle non-integer displacements.

\subsection{Signal interpolation}

Let us express the desired interpolation point $t + v(t)$ as
the sum of its entire part $k$ and fractional part $s$:
\begin{equation}
    x(t + v(t)) = x(k + s), \text{ with } k \in \mathbb{Z} \text{ and } s \in \left[0, 1\right[
\end{equation}
For the sake of clarity, the case $k = 0$ is considered. Note that this is
without loss of generality as this case can always be achieved through a change
of index for the function $x$. The objective of the interpolation is thus to
derive a value for $x(s)$.
\newline

Interpolation follows a two-step process. First, an $N$-tap filter $\mathbf{h} = \begin{bmatrix} h_1, \ldots, h_N \end{bmatrix}$,
is derived based on the fractional displacement $s$. Then, this filter is
applied on the $N$ closest (integer) neighbors of $s$ to obtain the interpolated
value $x(s)$:

\begin{equation}
    x(s) = \sum_{i = 1}^{N} h_i\, x(-\tfrac{N}{2} + i).
    \label{eq:warp-applied}
\end{equation}
As in conventional codecs, the filter length $N$ is restricted to even values
to have an identical number of neighbors left and right of the interpolation
point $s$ as illustrated in Fig. \ref{fig:1d-interpol}.

\subsection{Deriving the interpolation filter}

In order to derive an $N$-tap polynomial interpolation filter $\filter$, an $N
\times N$ matrix $\polynomialMatrix$ is applied to to a vector $\mathbf{s}$
based on the fractional displacement $s$:
\begin{equation}
    \filter = \polynomialMatrix\mathbf{s}, \text{ with } \mathbf{s}
    = \begin{bmatrix} s^0 & s^1 & \cdots & s^{N-1} \end{bmatrix}^\mathsf{T}.
    \label{eq:polynomial}
\end{equation}
Usual polynomial interpolations include the linear ($N=2$) and cubic
($N=4$) interpolations. Following Leguay et al. \cite{coolchic-2-leguay} to
derive their respective matrices $\polynomialMatrix$ yields:
\begin{equation}
    \small
    \polynomialMatrix_{lin} =
    \begin{bmatrix}
        \begin{array}{rr}
        1 & -1 \\
        0 & 1 \\
        \end{array}
    \end{bmatrix},\
        \polynomialMatrix_{cub} =
    \frac{1}{4}
    \begin{bmatrix}
    \begin{array}{rrrr}
        0 & -3 & 6  & -3 \\
        4 & 0  & -9 & 5 \\
        0 & 3  & 6  & -5 \\
        0 & 0  & -3 & 3 \\
    \end{array}
    \end{bmatrix}
    .
\end{equation}

Conventional codecs (HEVC, VVC, ECM) rely on a windowed sinc function to
generate the filter \cite{ecm-interpolation-samuelsson}. It is parameterized by
the fractional displacement $s$ and the filter length $N$:
\def\neighbors{\kappa}
\begin{equation}
    % h_i = \cos\left(\frac{\pi}{N} \Delta_i \right)\, \mathrm{sinc}(\Delta_i),\
    % \Delta_i = s - (- \frac{N}{2} + i).
    h_i = \cos\left(\frac{(s - \neighbors_i)\pi }{N}\right)\, \mathrm{sinc}(s - \neighbors_i),\
    \neighbors_i = -\frac{N}{2} + i.
    \label{eq:vvc-warp}
\end{equation}

\subsection{Complexity}

There are two sources of complexity in the interpolation process. Filter
 coefficients must be derived from the fraction displacement, using eq
 \eqref{eq:polynomial} or \eqref{eq:vvc-warp}. Then the filter is applied
 following eq. \eqref{eq:warp-applied}. However, the computation of the filter
 coefficients can be removed by restricting the fractional displacement $x$ to a
 finite number of possible values, allowing to pre-compute and store all
 the possible filters. Consequently, the number of multiplications required to
 perform the one-dimensional warping with an $N$-tap interpolation filter is:
\begin{equation}
    \mathcal{C}_{1d}(N) = N \text{ MAC / pixel}.
    \label{eq:warp-1d-complexity}
\end{equation}

\section{Two-dimensional motion compensation}

\subsection{Interpolation in two dimensions}

Let us represent a video frame and its corresponding motion field as
two-dimensional signals $x(c, r)$ and $v(c,r)$ with $c$ indexing the columns and
$r$ the rows. Using the re-indexation trick from the one-dimensional case, the
value of the shifted frame $y(c, r)$ is obtained by interpolating the input
signal at the fractional position $(s_c, s_r)$:
\begin{equation}
    y(c, r) = \mathrm{warp}(x, v)(c, r) = x(s_c, s_r),\, s_c, s_r \in \left[0, 1\right[.
\end{equation}

Two-dimensional interpolation with an $N$-tap filter requires $N^2$ neighboring
values:~$N$ neighboring lines, each composed of $N$ values. The interpolation
then proceeds in two steps, applying the displacement along both dimensions
successively following the one-dimensional method, as presented in Fig.
\ref{fig:2d-interpol}.
\newline

First, the filter coefficients $\filter^{(c)}$ are derived based on the column
displacement $s_c$ as in the one-dimensional case. The filter $\filter^{(c)}$ is
used on each of the $N$ neighboring lines separately, applying the horizontal
displacement $s_c$ to each of them and generating $N$ horizontally displaced
samples:
\begin{align}
    x(s_c, j) = &\sum_{i=1}^{N}h^{(c)}_i x(- \tfrac{N}{2} + i, j), \\
    \text{with } j \text{ the line index, }& j \in \left\{-\tfrac{N}{2} + 1, \ldots, -\tfrac{N}{2} + N \right\}. \nonumber
\end{align}
Then the interpolation is applied on the rows. To do so, new filter coefficients
$\filter_{r}$ are obtained from the vertical displacement $s_r$. This filter is
then applied to the horizontally displaced samples to obtain the value $x(s_c,
s_r)$:
\begin{equation}
    x(s_c, s_r) = \sum_{j=1}^{N}h^{(r)}_j x(s_c, j).
\end{equation}

\subsection{Complexity}

A two-dimensional warping with an $N$-tap filter requires us to perform $N + 1$
one-dimensional warpings per pixel. As such, the complexity grows quadratically
with the filter length:
\begin{equation}
    \mathcal{C}_{2d}(N) = (N + 1)\, \mathcal{C}_{1d}(N) = N^2 + N \text{ MAC / pixel}.
    \label{eq:warp-2d-per-pixel}
\end{equation}
% This complexity corresponds to the warping of a single pixel in a one-channel
% image. Modern codecs often warp a pair of 3-channel images (bidirectional
% prediction) with a filter length of $N = 8$ (\textit{e.g.} HEVC and VVC). This
% would bring about an overall warping complexity of $2 \times 3 \times
% \mathcal{C}_{2d}(8) = 432$ multiplications per pixel.
% \newline

Warping complexity is lowered by constraining the displacement to be identical
on a block of $B\times B$ pixels. This allows the re-use some of the
intermediate column computations, reducing the complexity of the warping to:
\begin{equation}
    \mathcal{C}_{\text{2d-block}}(N, B) = \frac{N^2 - N}{B} + 2N \text{ MAC / pixel}.
    \label{eq:warp-2d-block}
\end{equation}

\begin{table}[t]
    \caption{Typical motion compensation parameters.}
    \centering
    \small
    \begin{tblr}{
        colspec={Q[l] Q[c] Q[c] Q[c]},
        % rowsep=4pt,
        row{1} = {font=\bfseries,darkgray!5,belowsep=0.25mm, c},
        row{2} = {font=\bfseries,darkgray!5,abovesep=0.25mm, c},
        cell{1}{1}={c=1,r=2}{l,darkgray!5,font=\bfseries},
    }

    Codec       & Filter        & Motion            & Fractional \\
                & length $N$    &  block size $B$   & accuracy $\fractionalAccuracy$ \\
    \cmidrule[1pt,darkgray]{1-Z}
    HEVC \cite{overview-hevc-sullivan}   & 8           & 4 to 64              & 4                                  \\
    VVC \cite{overview-vvc-bross}        & 8           & 4 to 128              & 16                                   \\
    ECM \cite{ecm-16-seregin}            & 12          & 4 to 256              & 64                                   \\
    \cmidrule[1pt,darkgray]{1-Z}
    DCVC-FM \cite{dcvc-fm-li}           & 2                 & 1                     & Infinite                                  \\
    Cool-chic 4.0 \cite{coolchic-video-4-leguay} & 2                 & 1                     & Infinite                                  \\
    \end{tblr}
    \label{table:typical-value}
\end{table}

\section{Experiments}

\begin{figure}[t]
    \centering
    \includegraphics[width=\linewidth]{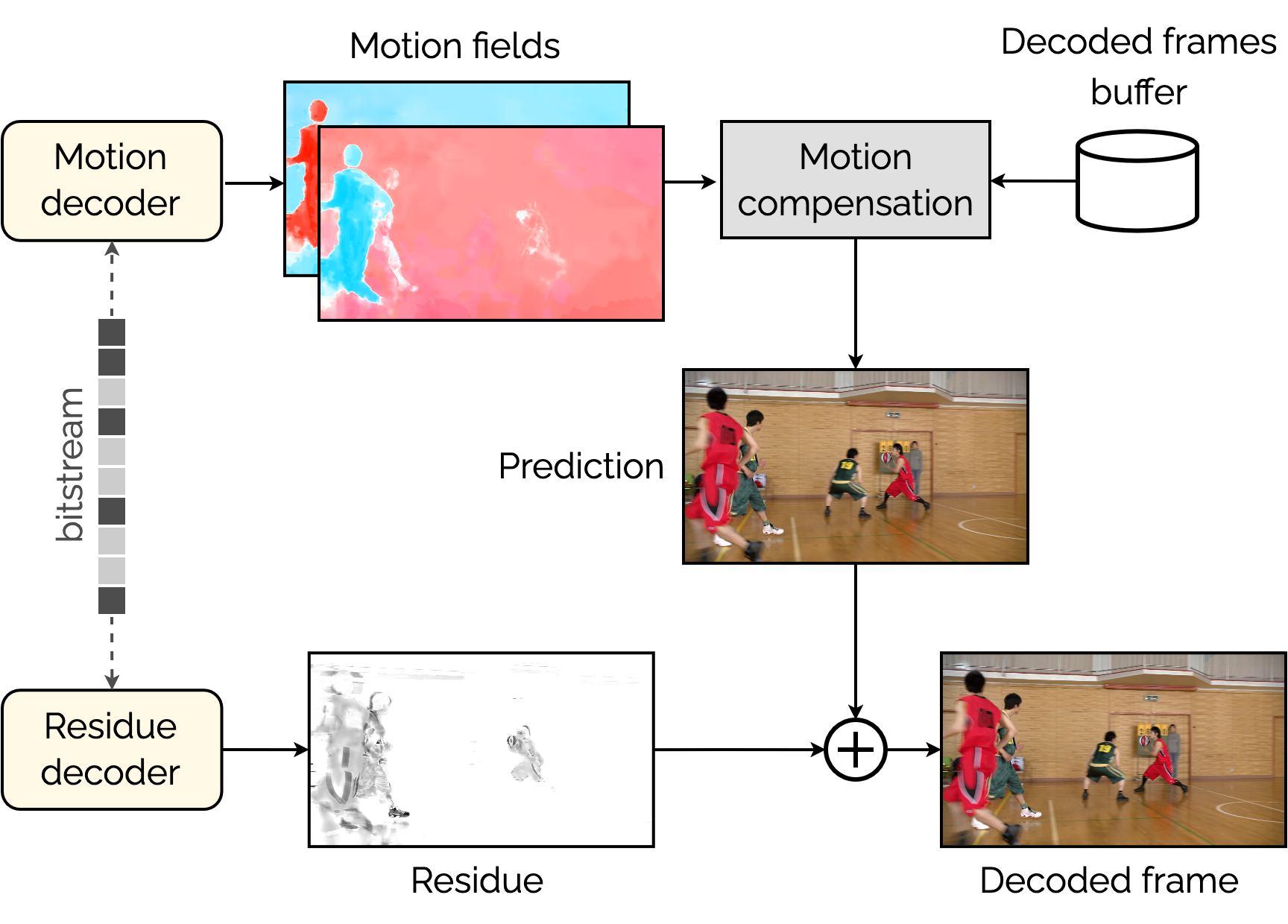}
    \caption{Decoding a B-frame with Cool-chic. Simplified diagram from Leguay
    et al. \cite{coolchic-video-4-leguay}.}
    \label{fig:cool-chic-decode}
\end{figure}

\subsection{Test conditions and baseline}

This section studies the complexity-performance trade-offs offered by the
equation \eqref{eq:warp-2d-block}. Namely, different interpolation filter
lengths $N$ and motion block sizes $B$ are investigated. The $N=2$ and $N=4$,
interpolation filters, corresponding respectively to bilinear and bicubic filters,
are obtained from the polynomial equation \eqref{eq:polynomial}. Longer
filters are derived following the conventional codecs sinc-based equation
\eqref{eq:vvc-warp}.

Table \ref{table:typical-value} presents the typical values of the motion
compensation parameters for different codecs. Conventional codecs leverage long
interpolation filters, maintaining a reasonable complexity through block-based
motion and finite fractional accuracy. Most learned codecs (\textit{e.g.} the
Cool-chic and DCVC series) rely on a simple bilinear interpolation filter,
applied pixel-wise with virtually infinite fractional accuracy. Moreover,
the DCVC family operates the motion compensation in a learned feature space,
alleviating some of the limitations of the bilinear filter at the cost of an
increased complexity. This is not the case for Cool-chic as it aims for low
complexity decoding.
\newline

The rate-distortion performance of each setting is evaluated on the first 33
frames of the JVET test sequences \cite{hevc-ctc-bossen} under a hierarchical
random access coding structure composed of an intra frame at each end and 31
B-frames in between. This study is carried out using the low complexity
overfitted codec Cool-chic 4.0 \cite{coolchic-video-4-leguay} as a baseline,
based on its open-source implementation \cite{coolchic-open-source-orange}.
The Cool-chic decoding process consists of three main steps, presented in Fig.
\ref{fig:cool-chic-decode}
\begin{enumerate}
    \item Decode one motion field per reference frame;
    \item Compute the temporal prediction through a bi-directional warping, in
    the YUV444 domain;
    \item Decode the residue and add it to the prediction.
\end{enumerate}

\subsection{Motion compensation complexity}

B-frames have 2 reference frames, with 3 channels each. Therefore, the motion
compensation complexity is 6 times the value mentioned in eq.
\ref{eq:warp-2d-block}. Table \ref{table:warp-complexity} presents the overall
motion compensation complexity for different motion block sizes $B$ and filter
lengths $N$. The quadratic growth of the complexity with respect to $N$ results
in an important complexity which is mitigated by increasing the motion block
size $B$.

Increasing the motion block sizes $B$ also reduces the number of motion vectors
composing the motion field. As such this decreases the overall motion decoder
complexity. This is illustrated in Table \ref{table:warp-complexity}, where
a motion block size $B=4$ reduces the complexity of the motion decoding to 34
multiplications per decoded pixel against 355 for the pixel-wise case ($B=1$).
\newline

Note that all the complexity mentioned in this paper only concerns the motion
part of the decoding process: motion decoding and motion compensation. Residue
decoding is not affected by the proposed changes and its complexity remains
fixed at 574 MAC per decoded pixel. Similarly, Intra frame decoding is kept
identical with a complexity of 1432 MAC per decoded pixel.

\begin{table}[t]
    \caption{Complexity (multiplications per decoded pixel) for the motion
        compensation and motion decoding with different filter sizes $N$ and
        block sizes $B$. Motion compensation complexity corresponds to
        $6\times\mathcal{C}_{\text{2d-block}}(N, B)$, see eq.
        \eqref{eq:warp-2d-block}.}
    \centering
    \small
    % \begin{tblrtikzbelow}
    % \draw[color=white,thick]
    % (h1-|v1) -- (h1-|v2) -- (h2-|v2)
    % -- (h2-|v3) -- (h3-|v3) -- (h3-|v4)
    % -- (h4-|v4) -- (h4-|v5) -- (h2-|v5)
    % -- (h2-|v6) -- (h1-|v6);
    % \end{tblrtikzbelow}%
    \begin{tblr}{
        colspec={Q[l] Q[l] Q[c] Q[c] Q[c] Q[c] Q[c]},
        row{1}={font=\bfseries},
        vspan=even,
        % Row name
        cell{4}{1}={c=1,r=3}{l,darkgray!5},
        cell{4}{2}={c=1,r=1}{l,darkgray!5},
        cell{5}{2}={c=1,r=1}{l,darkgray!5},
        cell{6}{2}={c=1,r=1}{l,darkgray!5},
        % Col name
        cell{1}{3}={c=4,r=1}{c,darkgray!5},
        cell{1}{7}={c=1,r=3}{c,darkgray!5},
        cell{2}{3}={c=1,r=1}{c,darkgray!5},
        cell{2}{4}={c=1,r=1}{c,darkgray!5},
        cell{2}{5}={c=1,r=1}{c,darkgray!5},
        cell{2}{6}={c=1,r=1}{c,darkgray!5},
        cell{3}{3}={c=1,r=1}{c,darkgray!5},
        cell{3}{4}={c=1,r=1}{c,darkgray!5},
        cell{3}{5}={c=1,r=1}{c,darkgray!5},
        cell{3}{6}={c=1,r=1}{c,darkgray!5},
        % cell{2}{7}={c=1,r=1}{c,darkgray!5},
        % % % Cells which are too complex
        % % vline{5}={3-3}{solid, fg=myred,1pt},
        % % vline{6}={4-6}{solid, fg=myred,1pt},
        % % hline{4}={5-5}{solid, fg=myred,1pt},
        % % hline{6}={6-Z}{solid, fg=myred,1pt},
        % % % I need to add some lines with the same color as the background
        % % vline{5}={1-2}{solid, fg=darkgray!5,1pt},
        % % vline{6}={1-2}{solid, fg=darkgray!5,1pt},
        % hline{4}={2-2}{solid, fg=darkgray!5,1pt},
        % Big dark gray lines
        % vline{2}={3-Z}{solid, fg=darkgray,0.25pt},
        vline{3}={1-Z}{solid, fg=darkgray,1pt},
        vline{7}={0-Z}{solid, fg=darkgray,1pt},
        row{3} = {belowsep=0mm, abovesep=0mm},
        }
        & & Motion compensation filter &  & & & {Motion \\decoding}\\
        % \cmidrule[0.25pt,darkgray]{3-6}
        & & $N=2$ & $N=4$  & $N=8$ & $N=12$ & \\
        & & bilinear & bicubic  & sinc & sinc & \\
        \cmidrule[1pt,darkgray]{1-7}
        \rotatebox[origin=c]{90}{\textbf{Motion block size} $B$} & 1 & 36 & 120 & 432 & 936 & 355\\
                                                        & 4 & 27 & 66 & 180 & 342 & 34 \\
                                                        & 8 & 26 & 57  & 138 & 243 & 13 \\
    % \hline[1pt,darkgray]
    \end{tblr}
    \label{table:warp-complexity}
\end{table}

\subsection{Performance-complexity trade-off}

\begin{figure}[t]
    \centering
    \begin{tikzpicture}
    \begin{axis}[
        xlabel={Decoding complexity [MAC / pixel] $\downarrow$},
        ylabel={BD-rate vs. baseline [\%] $\downarrow$},
        grid=major,
        minor y tick num=3,
        minor x tick num=4,
        legend pos=north east,
        % log ticks with fixed point,
        xtick distance=500,
        ytick distance=4,
        xmin=0, xmax=1500,
        ymin=-16, ymax=4,
        title={.},
        nodes near coords,
        point meta=explicit symbolic,
        nodes near coords align=horizontal,
    ]

    \addplot[
        % only marks,
        mark=*,
        color=myred,
        mark size=2.5pt,
        thick,
        smooth,
        nodes near coords style={
            anchor=south west,
        },
    ]
    coordinates {
        (391, 0) [$2$]
        (475, -10.13) [$4$]
        (787, -12.21666667) [$8$]
        (1291, -12.92666667) [$12$]
    };
    \addlegendentry{$B = 1$}

    \addplot[
        only marks,
        mark=triangle*,
        color=mygreen,
        mark size=2.5pt,
        thick,
        smooth,
        nodes near coords style={
            anchor=south,
            xshift=3pt,
            yshift=0.5pt
        },
    ]
    coordinates {
        (61, 1.846666667) [$2$]
        (100, -8.86) [$4$]
        (214, -10.5) [$8$]
        (376, -11.18) [$12$]
    };
    \addlegendentry{$B = 4$}

    \addplot[
        % only marks,
        mark=square*,
        color=myblue,
        mark size=2.5pt,
        thick,
        smooth,
        nodes near coords style={
            anchor=north east,
        },
    ]
    coordinates {
        (39, 2.533333333) [$2$]
        (70, -8.1) [$4$]
        (151, -10.1) [$8$]
        (256, -10.74) [$12$]
    };
    \addlegendentry{$B = 8$}

    % Add a specific point with a label
    \addplot[
      only marks,
      mark=o,
      mark size=5pt,
      ultra thick,
    ] coordinates {(391, 0)}
    node[pos=4, above, yshift=0.3cm] {Baseline};
    \end{axis}
    \end{tikzpicture}
    \caption{Compression performance as a function of the complexity (motion
    decoding and compensation). Average BD-rate on the first 33 frames of JVET
    classes B, C and E \cite{hevc-ctc-bossen} Each curve corresponds to a motion
    block size $B$ and is obtained by varying the filter length $N$ whose value
    is specified next to the data points.}
    \label{fig:performance-complexity}
\end{figure}
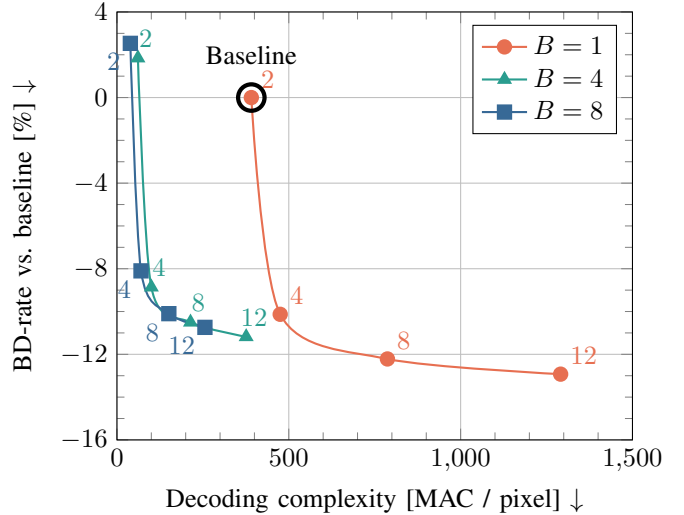

Figure \ref{fig:performance-complexity} presents the performance-complexity
trade-off offered by different warping parameters $(N, B)$. As presented in
Table \ref{table:warp-complexity}, increasing the filter length results in
higher decoding complexity. In turn, this increase in decoding complexity can be
alleviated by increasing the motion block size $B$. Using the longest
$N=12$-tap filter alongside a pixel-wise motion precision ($B=1$) gives the
biggest compression performance improvement. It offers a BD-rate of -13~\%
compared to the baseline ($N=2$, $B=1$). However, this improved compression
efficiency comes at the cost of an increase in the decoding complexity which
goes from 391 MAC / pixel to 1291 MAC / pixel.
\newline

When applying the motion compensation to block of pixels, increasing $B$ to 4 or 8
allows us to significantly decrease the overall decoding complexity while keeping
most of the compression gains. We propose to use an $N=8$-tap filter with motion
block size $B=4$, which results in a BD-rate of $-10.5$~\% and decreases the
decoding complexity from 391 MAC per pixel to 214 MAC per pixel.
\newline

Finally, Figure \ref{fig:rd-curve-jvet_b} presents the rate-distortion
performance of the proposed improved warping compared to Cool-chic 4.0
(baseline) and several conventional codecs. The improved warping reduces the
BD-rate of Cool-chic against HM from 41.9~\% to 14.3\%, outperforming
x265-medium and confirming the relevance of our contributions. We claim that
the $N=8$-tap sinc-based filter used alongside a motion block size $B=4$ should
replace the bilinear filter applied pixel-wise usually found within learned codec

\subsection{Quantized fractional displacement}

To ensure efficient decoder implementation, the filter coefficients have to be
pre-computed and stored. Consequently, it is required to limit the number of available
fractional displacement to a finite value $\fractionalAccuracy$. This is
achieved by quantizing the fractional displacement (vertical or horizontal) $s$ into:
\begin{equation}
    \hat{s} = \frac{\mathtt{round}(\fractionalAccuracy s)}{\fractionalAccuracy}\,.
    \label{eq:quantized-flow}
\end{equation}

In the case of Cool-chic (and all neural-based codec), gradient descent is used
to optimize the model. In this context it is more straightforward to
\textit{not} quantize the motion field during training as continuous values are
required for gradient-based optimization. This section shows that the motion
field can be quantized only at inference-time, to a small number of possible values
$\fractionalAccuracy$, with virtually no degradation in compression performance.
\newline

\begin{figure}[t]
    \centering
    \begin{tikzpicture}
    \begin{axis}[
        xlabel={Number of possible fractional displacement $\fractionalAccuracy$},
        ylabel={BD-rate vs. infinite motion accuracy [\%] $\downarrow$},
        xmode=log,
        log basis x=2,
        ymin=0, ymax=2.5,
        grid=major,
        minor y tick num=1,
        legend pos=north west,
        log ticks with fixed point,
        xmin=8, xmax=128
    ]

    \addplot[
        % only marks,
        mark=*,
        color=myred,
        mark size=2.5pt,
        thick,
    ]
    coordinates {
        % (4, 6.58)
        (8, 2.1675)
        (16, 0.6325)
        (32, 0.185)
        (64, 0.0625)
        (128, 0.0175)
    };

    \end{axis}
    \end{tikzpicture}
    \caption{Impact of the fractional displacement quantization to $\fractionalAccuracy$ values
        compared to the infinite precision case ($\fractionalAccuracy = \infty$). Results
        obtained on the first 33 frames of the JVET test sequences, with a
        $N=8$-tap sinc-based filter and motion blocks of size $B=4$.
    }
    \label{fig:quant-flow}
\end{figure}
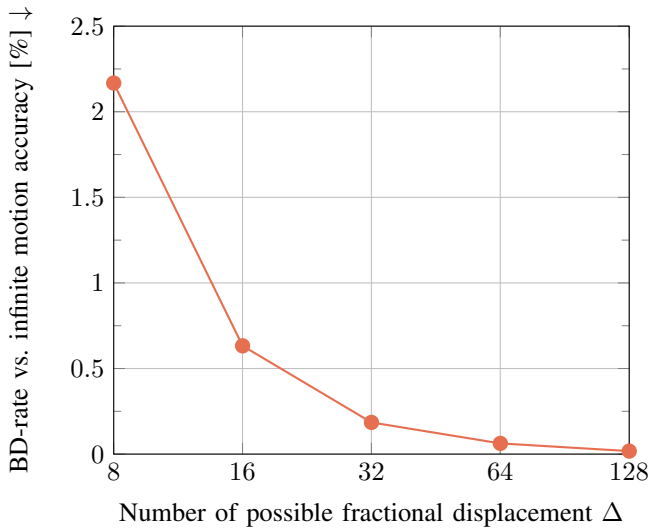

Figure \ref{fig:quant-flow} presents the degradation in compression performance
compared to the case where the fractional displacement is not quantized
($\fractionalAccuracy = \infty$). It shows that it is possible to reduce the
number of possible fractional displacements to a small number
($\fractionalAccuracy = 64$), while having a slight increase in BD-rate of only
0.06~\%. Consequently, we recommend to use $\fractionalAccuracy = 64$ values for
the fractional displacements alongside an $N=8$-tap interpolation filter as it
results in little to no degradation in the compression performance at the cost
of a minimal storage requirement ($8 \times 64 = 512$ pre-computed filter
coefficients).

\section{Conclusion}

This paper proposes to both improve the compression performance and reduce the
decoding complexity of learned video codecs, especially for the motion-related
coding steps. This is achieved by drawing inspiration from conventional codecs,
which leverage longer interpolation filters alongside block-based quantized
motion vectors.

We propose to use an 8-tap interpolation filtered derived from conventional
codecs as a replacement for the bilinear interpolation filter. To mitigate the
additional complexity, motion information is quantized and shared for blocks of
$4 \times 4$ pixels, instead of the usual pixel-wise motion information.
Additional complexity reduction is obtained by constraining the motion field to
a set of $64$ possible values, allowing us to pre-compute all filter coefficients,
at the cost of a marginal memory storage.
\newline

All these contributions are applied to the low complexity overfitted video codec
Cool-chic. They are evaluated against HEVC in a random access configuration,
where they allow to reduce Cool-chic BD-rate from 41.9~\% to 14.3~\%.
Additionally, the proposed contributions reduce the motion-related decoding
complexity from 391 to 214 multiplications per pixel. Beyond the Cool-chic
example, we suggest that these improvements could be beneficial for the entire
learned video compression community as a replacement for the usual bilinear
interpolation and pixel-wise motion information. Future work includes extending
this work to other state-of-the-art learned codecs \textit{e.g.} the DCVC
series \cite{dcvc-fm-li}.

% References should be produced using the bibtex program from suitable
% BiBTeX files (here: strings, refs, manuals). The IEEEbib.bst bibliography
% style file from IEEE produces unsorted bibliography list.
% -------------------------------------------------------------------------
\bibliographystyle{IEEEbib}
\bibliography{
    bibliography/ai,
    bibliography/autoencoder,
    bibliography/conventional,
    bibliography/coolchic,
    bibliography/dataset_metrics,
    bibliography/misc,
    bibliography/overfitted,
}
\end{document}